\documentclass[a4paper]{article}
\usepackage{array}
\usepackage{INTERSPEECH2022}
\usepackage{multirow}
\usepackage{subfigure}
\usepackage{stfloats}
\usepackage{graphicx}
 
\title{ADFF: Attention Based Deep Feature Fusion Approach for Music Emotion Recognition}
\name{Zi Huang$^1$, Shulei Ji$^1$, Zhilan Hu$^2$, Chuangjian Cai$^2$, Jing Luo$^1$, Xinyu Yang$^1$$^{\ast}$ \thanks{*Corresponding author} }
\address{
  $^1$School of Computer Science and Technology, Xi’an Jiaotong University\\
  $^2$Media Technology Institute, Huawei Technologies Co., Ltd.
}
\email{\{hz\_withChina, taylorji, luojingl\}@stu.xjtu.edu.cn,  yxyphd@mail.xjtu.edu.cn,\\ \{huzhilan, caichuangjian\}@huawei.com}

\begin{document}

\maketitle
\begin{abstract}
Music emotion recognition (MER), a sub-task of music information retrieval (MIR), has developed rapidly in recent years. However, the learning of affect-salient features remains a challenge. In this paper, we propose an end-to-end attention-based deep feature fusion (ADFF) approach for MER. Only taking log Mel-spectrogram as input, this method uses adapted VGGNet as spatial feature learning module (SFLM) to obtain spatial features across different levels. Then, these features are fed into squeeze-and-excitation (SE) attention-based temporal feature learning module (TFLM) to get multi-level emotion-related spatial-temporal features (ESTFs), which can discriminate emotions well in the final emotion space. In addition, a novel data processing is devised to cut the single-channel input into multi-channel to improve calculative efficiency while ensuring the quality of MER. Experiments show that our proposed method achieves 10.43\% and 4.82\% relative improvement of valence and arousal respectively on the \(R^2\) score compared to the state-of-the-art model, meanwhile, performs better on datasets with distinct scales and in multi-task learning.
\end{abstract}


\section{Introduction}

As an important entertainment for human beings, music has wide appeal. Studies indicate that people attach importance to music mainly because of the connotation and essential characteristics involved in emotion \cite{paper1}. More and more users retrieve music according to emotion in MIR systems, where MER plays a crucial role \cite{paper2,paper3}. However, the abstraction of emotion makes it difficult to analyze. Therefore, how to construct an effective MER method has attracted extensive attention.

The existing MER methods can be divided into two categories including classification and regression according to different emotion models. The former selects some emotional adjectives to classify music. However, the limited words cannot describe human emotions exactly \cite{paper4,paper5,paper6}. The latter uses the spatial position of emotion space to express human internal emotion. The two-dimensional valence-arousal (V-A) emotion model proposed by Russell \cite{paper7} is one of the mainstream emotion models for regression tasks \cite{paper8,paper9,paper10,paper11}. It represents emotion by valence and arousal, which stand for the degrees of pleasantness and bodily activation, respectively. In this paper, we aim to use a certain point in the V-A emotion space to describe the whole emotion for clip.

Feature engineering and model designing are the common solutions in MER. The first one improves recognition performance by constructing the effective feature sets \cite{paper13,paper14,paper29}. However, it requires a lot of manpower to design efficient feature sets for different datasets and recognition targets. For this reason, many researchers hope that the model can autonomously learn the affect-salient features. In the past, traditional machine learning methods (SVR, RF, etc.) were used to recognize music emotion \cite{paper9,paper12}, but gradually eliminated due to low flexibility and poor generalization. In recent years, deep learning has made amazing achievements in MER. The convolutional long-short-term-memory deep network based models proposed in \cite{paper15,paper16} can adaptively learn affect-salient features in music.  But these methods are generally suitable for a certain length and do not consider the relations among different clips. Besides audio, \cite{paper17,paper18} also introduce lyrics and other information to conduct multi-modal learning. Unfortunately, the lack of high-quality multi-modal aligned dataset leads to a significant reduction in applicable scenarios. Moreover, some researchers try to obtain more information by manipulating audio to improve the ability of their models. For example, \cite{paper8} tries to separate multiple sound sources in music (vocals, bass, drums, etc.) to explore the influence of different musical elements on MER. \cite{paper28} fuses multi-scale features of different lengths to strengthen predicting performance. Nevertheless, these methods are not strictly end-to-end structures, which may lead to additional errors when data flows through various modules.

To improve the above problems, we propose an end-to-end attention-based deep feature fusion (ADFF) approach for MER. We first take log Mel-spectrogram as input, then use adapted VGGNet as spatial feature learning module (SFLM) to obtain low-to-high level spatial structures. After that, these spatial features turn into multi-level emotion-related spatial-temporal features (ESTFs) by squeeze-and-excitation (SE) attention-based temporal feature learning module (TFLM). Finally, the prediction module maps the fusion into the emotion space. A series of experiments on the PMEmo dataset \cite{paper26} demonstrate that the ADFF model achieves an \(R^2\) score of 0.4575 for valence and 0.6394 for arousal respectively, which is a relative improvement of 10.43\% and 4.82\% compared to the state-of-the-art approach. It should be noted that the prediction of valence is notoriously more challenging than arousal \cite{paper13}. Furthermore, extended experiments on datasets with distinct scales and in multi-task also show that our method can effectively learn the affect-salient features from music clips and complete various tasks in MER.  

Our contributions are as follows: (1) our proposed ADFF model for MER achieves a better performance than the state-of-the-art method; (2) we introduce SE attention that enhances the weight of emotion-related features to help our model work better; (3) we design a novel data processing to improve calculative efficiency of the model while ensuring the quality of MER. 
\section{Propsed Method}

In this section, we mainly describe the ADFF model which consists of three modules: data processing, multi-level spatial-temporal feature learning, and prediction module. 
\begin{figure*}[ht]
	\centering
	\includegraphics[width=0.9\linewidth]{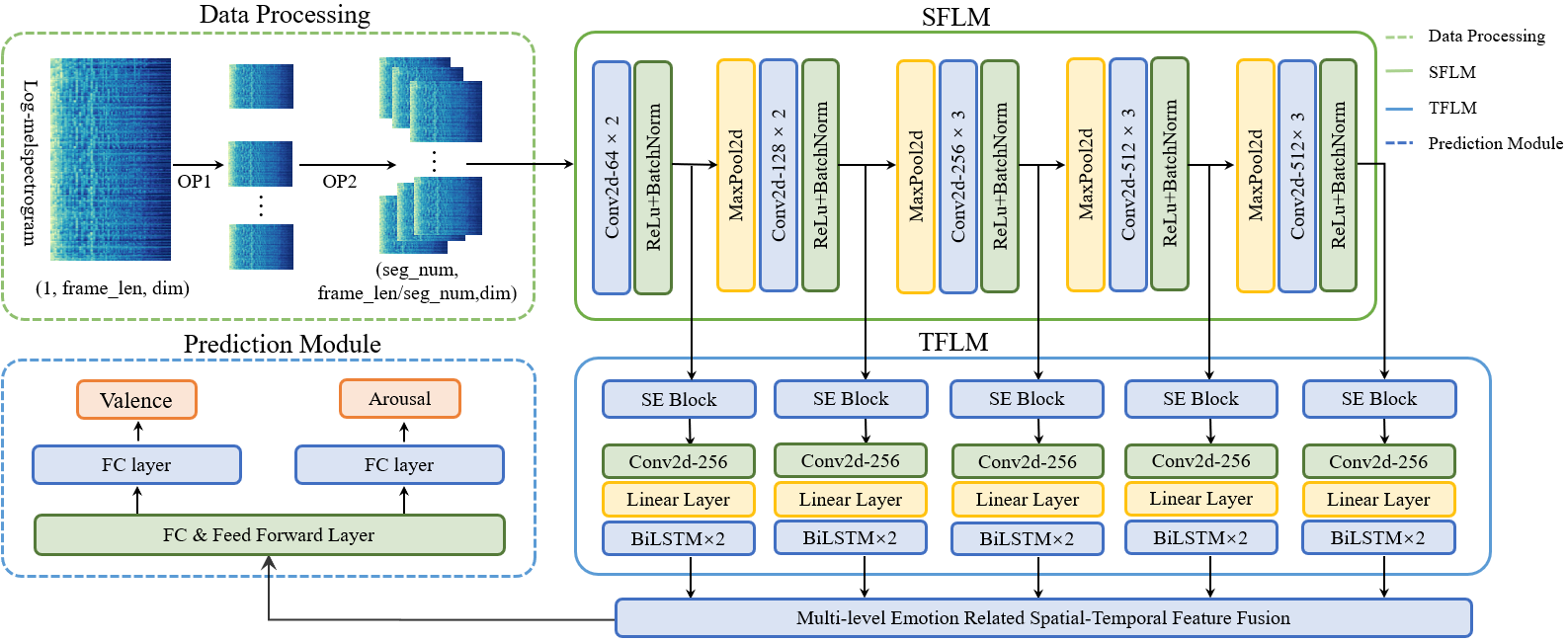}
	\caption{An overview of the ADFF model. OP1 means cutting the input into N segments, OP2 means stacking these segments into multi-channel. SFLM learns low-to-high level spatial features from input, and these multi-levels spatial features will be fed into the corresponding subnet of TFLM in each layer to obtain multi-level ESTFs. Finally, the fusion of ESTFs is mapped into the emotion space by the prediction module. You can choose the number of the last FC layer in prediction module for single-task or multi-task.}
	\label{fig:proposed_model}
\end{figure*}
\subsection{Data processing}
In this work, we use log Mel-spectrogram as input, whose dimension is (1, \(frame\_len\), \(dim\)), where \(frame\_len\) and \(dim\) represent the frame length and the number of Mel bands, respectively. The designed data processing method sequentially cut the spectrogram into \(seg\_num\) parts in the time dimension, and each part will form a new channel and be stacked into an image with dimension of \(\left( seg\_num, frame\_len/seg\_num, dim \right) \), as shown in the data processing module in Figure 1. We believe this method can reduce the distance among different intervals in the input and the long-term dependence. Experiments also show that our data processing method can improve the model computing efficiency while ensuring the model recognition performance.

\subsection{Spatial-temporal feature learning}
Learning emotional features is of vital importance to the MER task. It is commonly thought that emotion is involved in spatial and temporal features. For this reason, we design a specific spatial-temporal feature learning module for emotion learning, as shown by SFLM and TFLM in Figure 1. 

\subsubsection{Spatial feature leaning}
The effectiveness of VGGNet in image processing has been widely validated \cite{paper22}. In recent years, with the increasing attention of MER, some researchers have applied the VGGNet structure in it \cite{paper19,paper21,paper23,paper24}. However, they only explored their methods based on pipeline architecture in classification tasks. In this paper, we introduce the convolution subnetwork of VGGNet-16 as the spatial feature learning module (SFLM), which can be divided into 5 levels, and reform it by adding Relu and Batchnorm structures at the end of each level to get stronger and more robust spatial features. This adaptation makes SFLM more suitable for regression tasks. Suppose the N-th level of SFLM can be represented as \(SFLM_N\), and its output is \(S_N\):
\begin{equation}
	\begin{split}
	\left\{ \begin{array}{l}
		S_1=SFLM_1\left( input \right) \ ,\ if\ N=1\\
		S_N=SFLM_{N-1}\left( S_{N-1} \right) ,\ otherwise\\
	\end{array} \right. 
	\end{split}
	\label{eq1}
\end{equation}
where \(H_N\), \(W_N\), \(C_N\) represent the height, width and the number of channels of \(S_N\), respectively. If we divide \(S_N\) according to the number of channels, it also can be expressed as \(S_N=\left[ S_{N}^{1},S_{N}^{2},...,S_{N}^{C_N} \right] \). 
\subsubsection{SE attention-based temporal feature learning}
As a kind of temporal information, the music contains emotion changing over time. To capture temporally related emotions, we employ TFLM consisting of SE Attention and Bi-LSTM to learn temporal features. The SE attention is used to obtain the importance of different channels \cite{paper25}, which alleviates the problem of focusing on temporal structure only in traditional attention mechanisms. During the learning process, the input of a single channel is transformed into multi-channel through our data processing, which will strengthen the temporal correlation among distinct channels. In this case, the SE block can suppress the features that do not contribute much to emotion by learning the importance of different channels. Figure 2 shows the transformation of SE block, \(F_{sq}\left( \cdot \right) \) represents the squeeze operation.
\begin{equation}
	\begin{split}
		Z^{C_N}=F_{sq}\left( S_{N}^{C_N} \right) =\frac{1}{H_N\times W_N}\sum_{i=1}^{H_N}{\sum_{j=1}^{W_N}{S_{N}^{C_N}\left( i,j \right)}}\\
	\end{split}
	\label{eq2}
\end{equation}
By (\ref{eq2}) we get \(Z_N=\left[ Z_{N}^{1},Z_{N}^{2},...,Z_{N}^{C_N} \right] \), which is the global information embedding of the N-th level. To obtain the importance among the channels of \(S_N\), \(Z_N\) needs to go through an excitation transform yet.

\begin{figure}[ht]
	\centering
	\includegraphics[width=\linewidth]{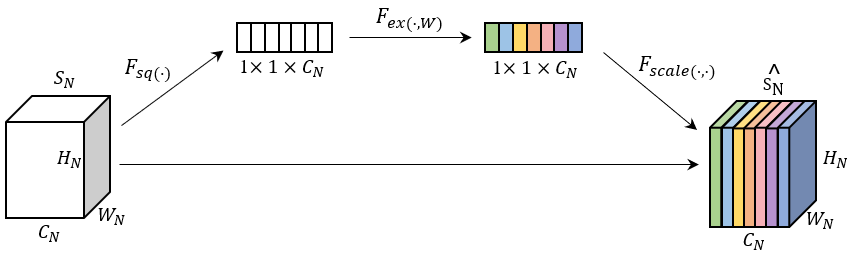}
	\caption{SE block of TFLM\protect\footnotemark.}
	\label{fig:SE block}
\end{figure}

\begin{equation}
	\begin{split}
	A_N=F_{ex}\left( Z_N,W \right) 
	\end{split}
	\label{eq3}
\end{equation}
where \(W_N\) and \(A_N\) represent learnable weight matrix and attention matrix. Then we get the weighted feature map \(\widehat{S_N}\) by rescaling \(S_N\) with \(A_N\):
\footnotetext{This figure is adapted from figure 1 of \cite{paper25}}
\begin{equation}
	\begin{split}
		\widehat{S_{N}^{C_N}}=F_{scale}\left( S_{N}^{C_N},A_{N}^{C_N} \right) =A_{N}^{C_N}\cdot S_{N}^{C_N}
	\end{split}
	\label{eq4}
\end{equation}
By (4), we get \(\widehat{S_N}=\left[\widehat{S_{N}^{1}},\widehat{S_{N}^{2}},...,\widehat{S_{N}^{C_N}} \right] \), which will be transformed to \(E_N\) (referred to ESTF) by 2-layer Bi-LSTM to complete temporal structure learning.
\subsection{Fusion strategy and emotion prediction}
Deep learning network can learn high-level abstract emotion-related features for target tasks, but some low-level features useful may be lost in this process \cite{paper20}. To make full use of all the multi-layer emotion-related information, we simply concatenate all the ESTF as  \(E=\left[ E_1,E_2,...,E_5 \right] \) to preserve the internal structure of ESTF from different levels maximumly. After that, \(E\) is fed into the prediction module to map into the emotion space. As shown in Figure.1, the prediction module is composed of several Fully Connected (FC) Layer and a Feedforward (FF) Layer. The number of the last FC layer is 1 or 2 according to single-task or multi-task, which maps the activation of FF layer to target output.

\section{Experiments and Results}
\subsection{Datasets}
The PMEmo dataset contains 794 chorus clips (provided by audio) of popular songs and their corresponding valence-arousal annotations. 457 annotators from different countries and majors are invited to annotate this dataset, and each chorus received at least 10 annotations. To compare our model with existing methods and make full use of the dataset, we split the dataset into two sizes. And for each size, we have 6 different lengths of input to explore the performance of the ADFF model while dealing with various segments, that is, \(seg\_lens\)=[5, 10, 15, 20, 25, 30].
\begin{itemize}
\item {\bfseries Simple Datasets}. Each chorus will be cut into a specified fixed length randomly. Namely, each chorus and its corresponding annotation appear only once in a simple dataset, which is consistent with \cite{paper8,paper11}.
\item {\bfseries Full Datasets}. To make the most of the PMEmo dataset, we cut each chorus into several segments with specified lengths in time order. If the length of the last segment is less than \(seg\_len/2\), it will be dropped. Otherwise, it will be extended forward until satisfying the demand. All segments cut out from the same chorus share the same static annotation.
\end{itemize}

Note that no matter which dataset is selected, if the original length of the chorus is less than specified \(seg\_len\), the audio will be padding with zero first to ensure a fixed length. And all the valence and arousal annotations are linearly scaled to [-1, 1] to improve the robustness of the model.
\subsection{Experimental details}
We use the \(R^2\) score and root-mean-squared error (RMSE) in the regression task, and accuracy in the classification task as the evaluation metrics. The \(R^2\) score ranges from negative infinity to 1, and the larger value is better. On the contrary, the smaller the RMSE is, the better. For a fairer comparison and to avoid accidental errors, we take the mean result of 5-fold cross-validation as the final result following \cite{paper8,paper11}. The input log Mel-spectrogram is extracted by librosa 0.7.2 tool \cite{paper27}, with Mel bands of 128, a sampling rate of 44.1KHZ, window size and hop size of 60ms and 10ms respectively. Moreover, we use Adam optimizer for training, with a decay weight of 1e-5, learning rate of 1e-5, training epoch of 200, and batch size of 32. The decay steps are [20, 45, 80, 110, 140, 170].

\subsection{Experimental results}
\subsubsection{Method comparison and ablation analysis}
To verify the advancement of our proposed model and explore the role of different modules, we choose EmoMucs \cite{paper8}, MLEM \cite{paper11}, and two variants of the ADFF model for comparison. EmoMucs recognizes music emotion based on the source separation algorithm and announced it as the most advanced method. MLEM designes a useful method to debug the biased MER model to perform better. The two variants of the ADFF model drop SE-block and TFLM, respectively. Note that EmoMucs and MLEM both experiment on the simple dataset with the input length of 20 seconds, and we compare the best performance of each model in the same condition.

\begin{table}[t]
	\caption{The performance of different models on simple dataset with \(seg\_len\) of 20 seconds}
	\label{tab:different_model}
	\centering
	\scalebox{0.68}{
		\begin{tabular}{c c c c c}
			\toprule
			\multicolumn{1}{c}{\multirow{2}{*}{Model}}&
			\multicolumn{2}{c}{Arousal}&
			\multicolumn{2}{c}{Valence}\cr 
			\cmidrule(lr){2-3} \cmidrule(lr){4-5}
		   &RMSE&R2&RMSE&R2\cr 
			\midrule
			$EmoMucs$ & $0.2285$  & $0.6100$ & $0.2466$ & $0.4143$\\
			$MLEM$ & $0.2500\pm0.03$  & $0.6000\pm0.10$ & $0.3100\pm0.04$ & $0.4000\pm0.14$\\
			\midrule
			
			$ADFF$ & \pmb{$0.2213\pm0.01$}  &\pmb{ $0.6394\pm0.02$} & \pmb{$0.2379\pm0.03$} & \pmb{$0.4575\pm0.08$}\\
			$w/o\ \ SE$ & $0.2253\pm0.01$  & $0.6239\pm0.05$ & $0.2429\pm0.03$ & $0.4332\pm0.09$\\
			$w/o\ \ TFLM$& $0.2228\pm0.01$  & $0.6316\pm0.05$ & $0.2469\pm0.02$ & $0.4155\pm0.06$\\
			\bottomrule
			
		\end{tabular}
	}
\end{table}

As shown in Table 1, our proposed ADFF model has obvious advantages over the others, which get lower RMSE and higher \(R^2\) scores. Particularly, the \(R^2\) score of valence and arousal increase relatively 10.43\% and 4.82\% than EmoMucs, and note that valence is more challenging to predict than arousal. If taking the deviation into account, our method is more stable than MLEM. Moreover, the ADFF model belongs to a strictly end-to-end architecture, but EmoMucs and MLEM both need source separation algorithms to process the original audio before training, which may introduce more errors.

In addition, ablation experiments show that achieving high-performance prediction for arousal doesn't need a complicated model, this conclusion is consistent with \cite{paper30}. And compared with the two variants, the ADFF model has a relative improvement of 5.61\% and 10.11\% in \(R^2\) score for predicting valence, which means SE block enhances the weight of emotion-related features and TFLM further improves the emotion capture ability of model. 

\subsubsection{The influence of data processing}
To explore the influence of our data processing on the model, we compared the results of ADFF on simple datasets of 20 seconds when \(seg\_num\) changes in [1, 2, 4, 6, 8, 10, 12, 14, 16].

\begin{figure}[th]
	\centering
	\includegraphics[width=\linewidth]{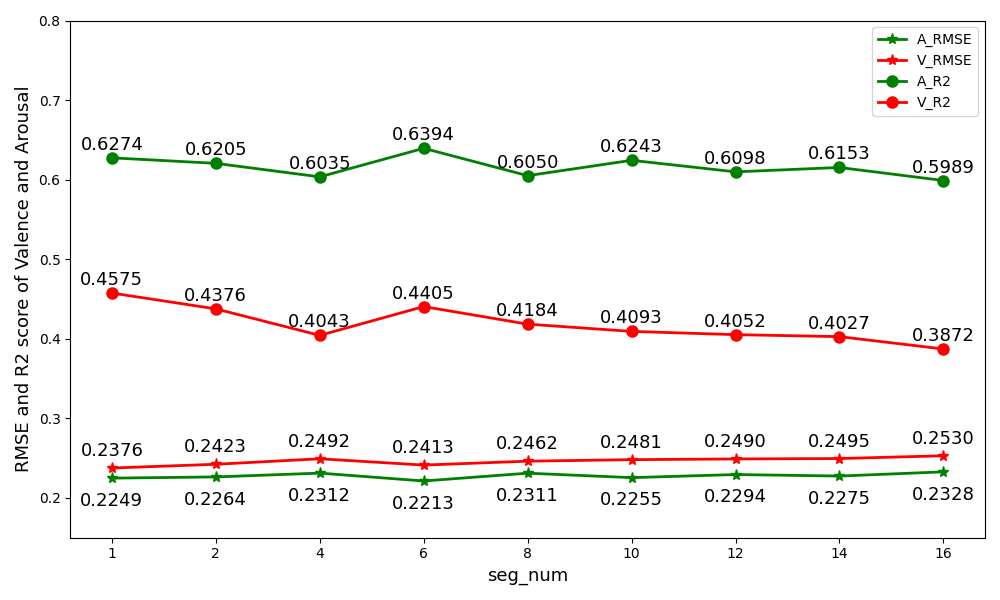}
	\caption{The \(R^2\) score and RMSE of Valence (V) and Arousal (A) with different \(seg\_num\) on simple dataset of 20 seconds.}
	\label{fig:differentSeg}
\end{figure}

\begin{table}[th]
	\caption{The cost time of different \(seg\_num\)}
	\label{tab:timeCost}
	\centering
	\scalebox{0.68}{
	\begin{tabular}{ cccccccccc }
		\toprule
		\multicolumn{1}{c}{\(seg\_num\)} & 
		\multicolumn{1}{c}{1} & 
		\multicolumn{1}{c}{2} & 
		\multicolumn{1}{c}{4} & 
		\multicolumn{1}{c}{6} & 
		\multicolumn{1}{c}{8} & 
		\multicolumn{1}{c}{10} & 
		\multicolumn{1}{c}{12} & 
		\multicolumn{1}{c}{14} & 
		\multicolumn{1}{c}{16} \\
		
		\midrule
$time(h)$ & $4.99$ & $1.52$ & $0.76$ &$0.51$ &$0.40$ & $0.28$ & $0.29$ & $0.27$ &$0.25$       \\

		\bottomrule
	\end{tabular}
}
\end{table}

Figure 3 shows that \(seg\_num\) has different optimal values for various tasks. The best and the second-best results achieve when \(seg\_num\) is 1 and 6 for valence, but for arousal, the \(seg\_num\) becomes 6 and 1. If considering the metric only, it's a good choice to choose whichever \(seg\_len\) of 1 or 6 for input with a length of 20 seconds. Furthermore, we observe that along with the growth of \(seg\_num\), the time cost will decrease clearly as shown in Table 2. However, if we choose a too large \(seg\_num\), the recognition performance of the model will reduce. We conjecture the reason is that the feature represented by each channel is too short to contain enough emotional information. Balancing the time cost and performance benefit, we choose \(seg\_num\) of 6 in the following experiments.

\subsubsection{The performance of different lengths}
Previous studies have shown that too long or too short input will damage the performance of MER \cite{paper16,paper31}. Exploring the influence of input with different lengths on the model can help to select the optimal input length and maximize the comprehensive benefits of recognition performance and speed. Therefore, we explore the performance of the proposed model when input length changes. In addition, we also explore the influence of data size on the performance of recognition.

\begin{table}[t]
	\caption{The performance of different lengths on simple and full datasets with \(seg\_num\) of 6.}
	\label{tab:table3}
	\scalebox{0.68}{
		\begin{tabular}{c c c c c c c c c}
			\toprule
			\multicolumn{1}{c}{}&
			\multicolumn{4}{c}{Simple Datasets}&
			\multicolumn{4}{c}{Full Datasets}\cr
		
			\cmidrule(lr){2-5} \cmidrule(lr){6-9} 
				\multicolumn{1}{c}{\multirow{2}{*}{}}&
				\multicolumn{2}{c}{Arousal}&
				\multicolumn{2}{c}{Valence}&
				\multicolumn{2}{c}{Arousal}&
				\multicolumn{2}{c}{Valence}\cr 
				\cmidrule(lr){2-3} 	\cmidrule(lr){4-5}	\cmidrule(lr){6-7} \cmidrule(lr){8-9}
				&RMSE&R2&RMSE&R2& RMSE&R2&RMSE&R2\cr 
				\midrule
			$5$ & $0.2350$  & $0.5914$ & $0.2532$ & $0.3850$ & $0.2262$  & $0.6203$ & $0.2337$ & $0.5013$\\
			$10$ & $0.2259$  & $0.6233$ & $0.2556$ & $0.3740$ & $0.2186$  & $0.6441$ & \pmb{$0.2312$} & \pmb{$0.5083$}\\
			$15$ & $0.2253$  & $0.6256$ & $0.2573$ & $0.3651$ & $0.2182$  & $0.6472$ & $0.2359$ & $0.4864$\\
			$20$ & \pmb{$0.2213$}  & \pmb{$0.6394$} & \pmb{$0.2413$} & \pmb{$0.4405$} & \pmb{$0.2160$}  & \pmb{$0.6545$} & $0.2378$ & $0.4785$\\
			$25$ & $0.2281$  & $0.6121$ & $0.2552$ & $0.3728$ & $0.2192$  & $0.6426$ & $0.2460$ & $0.4477$\\
			$30$ & $0.2296$  & $0.6090$ & $0.2507$ & $0.3959$ & $0.2260$  & $0.6193$ & $0.2521$ & $0.4172$\\
			\bottomrule
			
		\end{tabular}
	}
\end{table}


As shown in Table 3, different input lengths will result in the discrepancy of recognition performance. On simple datasets, the model works best when \(seg\_len\) is 20. Note that input with different lengths should have respective optimal \(seg\_num\) parameters, and we have not investigated one by one here. On full datasets, the recognition performance for all lengths have been improved. The \(R^2\) score of the valence of 5 seconds and 10 seconds segments stand out particularly, which relative increased by 30.21\% and 35.91\% respectively than on simple datasets, even outpaced the performance of 20 seconds. This illustrates that the size of dataset will affect the performance of the model. Even if a clip is short, excellent recognition performance can also be obtained when the model is trained fully. The conclusion may help the existing music software to improve its emotion-related recommendation function.

\subsubsection{The performance of multi-task and classification task}

\begin{table}[t]
	\caption{The performance of multi-task and classification-task on full datasets with \(seg\_num\) of 6.}
	\label{tab:table4}
	\centering
	\scalebox{0.7}{
		\begin{tabular}{c c c c c c c c}
			\toprule
			\multicolumn{1}{c}{\multirow{2}{*}{}}&
			\multicolumn{2}{c}{Arousal}&
			\multicolumn{2}{c}{Valence}&
			\multicolumn{2}{c}{Two-category}&
			\multicolumn{1}{c}{\multirow{2}{*}{Four-category}}\cr 
			\cmidrule(lr){2-3} 	\cmidrule(lr){4-5}	\cmidrule(lr){6-7}
			&RMSE&R2&RMSE&R2& Arousal&Valence\cr 
			\midrule
			$5$ & $0.2255$  & $0.6227$ & $0.2351$ & $0.4955$ & $82.40\%$  & $81.00\%$ &  $71.48\%$\\
			$10$ &$0.2190$  & $0.6428$ & \pmb{$0.2332$} & \pmb{$0.5004$} & $83.12\%$  & \pmb{$81.29\%$} &  \pmb{$71.90\%$} \\
			$15$ &\pmb{$0.2181$}  & \pmb{$0.6477$} & $0.2407$ & $0.4647$ & $83.27\%$  & $80.27\%$ &  $70.84\%$\\
			$20$ &$0.2188$  & $0.6449$ & $0.2394$ & $0.4713$ & \pmb{$83.60\%$}  & $80.52\%$ &  $70.88\%$\\
			$25$ &$0.2214$  & $0.6345$ & $0.2501$ & $0.4313$ & $83.30\%$  & $80.67\%$ &  $70.70\%$\\
			$30$ &$0.2248$  & $0.6252$ & $0.2537$ & $0.4098$ & $82.35\%$  & $80.14\%$ &  $70.26\%$\\
			\bottomrule
	
		\end{tabular}
	}
\end{table}

We also explore the ability of the proposed model to multi-task. As shown in Table 4, when the model predicts valence and arousal simultaneously, the performance is not much different from single-task. But doing so can avoid the trouble of model designing or training separately for various tasks. And it also demonstrates that our proposed model has an excellent ability to learn affect-salient features for both valence and arousal. Besides, we have extended our experiments on two-category and four-category tasks according to the positive and negative values of valence and arousal. Obviously, four-category tasks are more challenging than two-category, and predicting valence is more difficult than arousal. But to our surprise, different lengths of inputs have no obvious effect on the classification tasks.

\section{Conclusion and Future work}
In this work, we propose an end-to-end attention-based deep feature fusion method for MER. The proposed model builds a bridge from affect-salient feature to emotion space effectively by using log Mel-spectrogram only. A series of experiments prove that the proposed model outperforms the state-of-the-art method in performance and robustness, and maintains high recognition quality on different input lengths, dataset sizes, or tasks. Until now, we have only explored on PMEmo dataset, in which English songs account for the vast majority. However, there are various genres and languages of music in reality. In future work, we will investigate the ability of our model on cross-language or cross-genre datasets, and try to introduce pre-training to further improve the effectiveness of our model.

\bibliographystyle{IEEEtran}

\bibliography{reference}


\end{document}